\documentclass{nature}

\usepackage[displaymath]{lineno}

\usepackage{lmodern}
\usepackage{amsmath}
\usepackage{amssymb}
\usepackage{graphicx}
\usepackage{hyperref}
\usepackage{xspace}
\usepackage{caption}
\usepackage{mathptmx}
\usepackage{mathrsfs}
\usepackage{pdfpages}

\newcommand{\sms}{\ensuremath{\mathbf{S}}\xspace}

\newcommand{\TC}{\ensuremath{T_{\mathrm{C}}}\xspace}

\newcommand{\criii}{\ensuremath{\mathrm{CrI}_3} \xspace}

\title{Quantum rescaling, domain metastability and hybrid domain-walls in two-dimensional CrI$_3$ magnets} 

\author{Dina Abdul Wahab$^{1}$, Mathias Augustin$^{1}$, Samuel Manas Valero$^{2}$, Wenjun Kuang$^{3}$, Sarah Jenkins$^4$, Eugenio Coronado$^{2}$, Irina V. Grigorieva$^{3}$, Ivan J. Vera-Marun$^{3}$, Efr\'en Navarro-Moratalla$^{2}$, Richard F. L. Evans$^4$, Kostya S. Novoselov$^{3,5,6}$ \& Elton J. G. Santos$^{7,\dagger}$}

\makeatletter
\let\saved@includegraphics\includegraphics
\AtBeginDocument{\let\includegraphics\saved@includegraphics}
\renewenvironment*{figure}{\@float{figure}}{\end@float}
\makeatother

\begin{document}

\maketitle

\begin{affiliations}
\item School of Mathematics and Physics, Queen's University Belfast, BT7 1NN, UK
\item Instituto de Ciencia Molecular, Universidad de Valencia, Calle Catedr\'atico Jos\'e Beltr\'an 2, 46980 Paterna, Spain 
\item School of Physics, University of Manchester, Oxford Road, Manchester, M13 9PL, UK
\item Department of Physics, The University of York, York, YO10 5DD, UK 
\item Department of Material Science \& Engineering, National University of Singapore, Block EA, 
9 Engineering Drive 1, 117575, Singapore  
\item Chongqing 2D Materials Institute, Liangjiang New Area, Chongqing 400714, China 
\item Institute for Condensed Matter Physics and Complex Systems, School of Physics and Astronomy, The University of Edinburgh, EH9 3FD, UK. \\ 
$^{\dagger}$Corresponding author: esantos@ed.ac.uk  
\end{affiliations}
 
\date{}

\begin{abstract}
Higher-order exchange interactions and quantum effects are widely 
known to play an important role in describing the properties 
of low-dimensional magnetic compounds. 
Here we identify the recently discovered 
two-dimensional (2D) van der Waals (vdW) CrI$_3$ as a 
quantum non-Heisenberg material with properties far beyond 
an Ising magnet as initially assumed. 
We find that biquadratic exchange interactions 
are essential to quantitatively describe the magnetism of CrI$_3$ 
but requiring quantum rescaling corrections to reproduce 
its thermal properties. The quantum nature of the heat bath represented 
by discrete electron-spin and phonon-spin scattering processes induced the formation 
of spin fluctuations in the low temperature regime. 
These fluctuations induce the formation 
of metastable magnetic domains evolving into a single 
macroscopic magnetization or even a monodomain 
over surface areas of a few micrometers. Such 
domains display hybrid characteristics of N\'eel and Bloch 
types with a narrow domain wall width in the range of 3-5 nm. 
Similar behaviour is expected for the majority of 2D vdW 
magnets where higher-order exchange interactions are appreciable.   
\end{abstract}

\noindent 

The rediscovery of magnetism in layered vdW systems\cite{deJonghbook}  
has sparked an increasing interest in the investigation of spin interactions at 
the ultimate limit of few atom thick 
materials\cite{firstCrI3,CrGeTe,Song1214,Klein1218,Ghazaryan:2018aa,Wang:2018aa,Fei:2018aa,Seyler:2018aa,Tian_2016,Guguchiaeaat3672}. 
With the advent of new techniques of isolation, 
manipulation, measurements and theoretical predictions, 
vdW magnets have become a playground for achieving 
the limit of magnetism in atomically thin crystals and 
unveil novel physical phenomena. Implementation
of 2D vdW magnets in real technologies 
however requires the description of their magnetic 
properties through an archetypal spin model such as Ising or Heisenberg. 
This gives a predictive indicator to what kind of behaviour is
expected if such a magnet could be probed experimentally. 
For instance, Ising is incompatible 
with the appearance of magnons or spin-waves since just 
two spin states (e.g. $\pm z$) are taken 
into account\cite{Skomski}. 
If atomic spins ${\bf S}_{i}=\mu_{s}{\bf s}_i$, 
where $\mu_{s}$ is the magnetic 
moment, need to precess to a different spatial 
orientation the Heisenberg model 
would give unrestricted values of ${\bf S}_{i}$ on the unit sphere 
surface $|{\bf s}_i|=1$ in order to minimize the exchange interaction energy. 
As dimensionality determines the 
stabilization of spin ordering differently to the bulk phase,
previously demonstrated for many low-dimensional 
nanostructures\cite{Slonczewski,Bode07,Rohart13}, 
higher-order exchange interactions beyond the 
Heisenberg or Ising models would be 
expected to play a key role in the magnetic properties 
of the magnetic layers. Of particular interest is CrI$_3$ 
where magnetization has been firstly measured using 
magneto-optical Kerr effect setup\cite{firstCrI3} 
at the limit of monolayer. Although CrI$_3$ has been treated as 
an Ising ferromagnet due to its large anisotropy, 
recent findings reporting the appearance of 
topological spin-excitations\cite{Pengcheng18}, 
temperature dependent magnons\cite{Liuyan18} and 
angle-dependent ferromagnetic resonance\cite{Hammel19} 
indicate that the magnetic properties of CrI$_3$ are far 
beyond Ising. 

Here we show that these puzzling features can be naturally reconciled 
with the inclusion of biquadratic (BQ) interactions 
in an extended Heisenberg framework including 
additional quantum rescaling corrections. 
Our starting point is the following spin Hamiltonian: 
\begin{linenomath}
\begin{equation}
\label{eq:hamiltonian}
{H} = - \sum_{ij} J_{ij} (\sms_i \cdot \sms_j) - \sum_{i,j} \lambda_{ij} S_i^z S_j^z - \sum_i D_i \left(S_i^z\right)^2 - \sum_{ij} K_{ij} \left(\sms_i \cdot \sms_j \right)^2  
\end{equation}
\end{linenomath}
%
where $\sms_i$ is the localized magnetic moment unit vector on Cr atomic sites $i$ which 
are coupled by pair-wise exchange interactions. $J_{ij}$ and $\lambda_{ij}$ are the 
isotropic and anisotropic bilinear (BL) exchanges, respectively, and $D_i$ is the 
onsite magnetic anisotropy. We used up to third-nearest neighbors on both $J_{ij}$ and $\lambda_{ij}$. 
The fourth term represents a biquadratic (BQ) exchange which occurs due to the 
hopping of more than one electron between two adjacent sites\cite{Slonczewski,Kartsev:2020aa}. 
Its strength is given by the constant $K_{ij}$, which is 
the simplest and most natural form of non-Heisenberg coupling. 
We can determine $K_{ij}$ by calculating the energetic variation of the spins $\sms_i$ 
at each Cr site at different rotation angle $\theta$ including spin-orbit 
coupling\cite{PhysRevB.90.014437, Nicola15,NOVAK2008} 
(Figure \ref{fig1}{\bf a}). See Supplementary Sections S1-S4 for details, 
and comparison with other models, such as Kitaev\cite{Kitaev06,Laurent18}.
%
%
It is noteworthy that $\lambda_{ij}$ and $K_{ij}$  in monolayer CrI$_3$ 
have close magnitudes but are slightly smaller than $J_{ij}$ (Supplementary Table S1). 
Indeed, in materials where the exchange is for 
some reason weak due to different processes, such as competition between 
ferromagnetic and antiferromagnetic exchange\cite{Nagaev_1982}, 
BQ exchange has a particularly strong influence as observed 
for several different compounds\cite{Wysocki:2011aa,Bode07,Harris63,Ashvin09,Iron-BQ}. 
This seems to be the case for most of the 
vdW magnets as recently demonstrated\cite{Kartsev:2020aa}. 
We can apply similar analysis to bulk CrI$_3$ which shows the 
same magnitude of BQ exchange for the intra-layer interactions but smaller for the 
interlayer counterparts (Supplementary Table S1).  
These results indicate that higher-order exchange processes involving two or more 
electrons are important in CrI$_{\rm 3}$ magnets despite of the dimensionality. 
Nevertheless, we focus on the effect of BQ exchange on the magnetic features of CrI$_3$
not considering higher order interactions, i.e. three-site spin interactions\cite{PhysRevB.101.024418}.

To simulate the temperature and dynamic properties of CrI$_3$ at a macroscopic level, 
we have implemented the BQ exchange interactions shown in Eq.\ref{eq:hamiltonian} 
within the Monte Carlo Metropolis algorithm\cite{Evans14} also taking into account 
contributions from the next-nearest neighbours. See Supplementary Sections S5 for details.  
In this Monte Carlo model we assume a classical spin vector $\sms_i$ on each atomic site $i$ 
of fixed length $\mu_{s}$ whose direction can vary freely in 3D space. 
The quantization vector for the spin is 
a local quantity which naturally includes the effects of 
local thermal spin fluctuations and magnon processes. 
This clearly separates classical and quantum contributions to the magnetic behaviour of CrI$_3$.
To analyse whether the Ising or the non-Heisenberg model (Eq.\ref{eq:hamiltonian}) provides the best 
description of the magnetic properties of bulk and monolayer CrI$_3$, we undertake a 
quantitative comparison between both models with the measurement of the 
magnetization $M$ versus temperature $T$ using first-principles parameters as input. 
We use the magneto-optical Kerr effect (MOKE) data 
extracted from Ref.\cite{firstCrI3} for monolayer CrI$_3$, 
and superconducting quantum interference device (SQUID) technique 
for bulk CrI$_3$ (Supplementary Sections S6 and S7). 
Figure~\ref{fig1}{\bf a-b} show the simulated temperature dependence of the 
magnetization of bulk and monolayer CrI$_3$ relative to the experimental data. 
It is clear that the Ising model grossly overestimates the measured Curie temperature ($T_C$) 
for both systems by several tens of Kelvins reaching high temperatures. 
Ising gives $T_C \sim$200 K and $T_{C} \sim$102 K 
for bulk and monolayer CrI$_3$, respectively, which is 
also in disagreement with previous experimental 
studies\cite{firstCrI3,Petrovic18,Lin18,Luo19}.  
This suggests that a single quantization 
axis where spins are allowed to take only two values 
parallel or anti-parallel to the surface is not accurate 
enough to represent the magnetic properties of CrI$_3$ magnets.
Conversely, the non-Heisenberg model gives a sound agreement with the measurements 
resulting in Curie temperatures of 44.4 K and 62.2 K for monolayer 
and bulk CrI$_3$, respectively.  
We have also checked whether other models  
can give a sound description of the magnetic properties of CrI$_{\rm 3}$. 
Namely, a Heisenberg model without the inclusion of 
BQ interactions\cite{cri3}, a Kitaev model\cite{Kitaev06,Laurent18}, 
and also the BQ model in Eq.\ref{eq:hamiltonian} including 
non-collinear spin-textures at the level of Dzyaloshinskii-Moriya 
interactions (DMI). See Supplementary Sections 4 and 8 for details.  
While DMI do not give any variation of $T_{C}$ relative to the initial 
BQ model, both the Heisenberg and Kitaev models 
significantly underestimate the magnitude of the 
critical temperatures by several tens of Kelvin's relative to the 
measurements ($T_{\rm C }^{Kitaev}=17.2$ K, $T_{\rm C }^{Heisenberg}=23 - 37.4$ K). 
These results suggested that BL models are insufficient to describe the 
magnetic features of CrI$_3$. Furthermore, the inclusion of BQ 
exchange has recently been observed in the 
description\cite{Kartsev:2020aa} of neutron scattering 
measurements on the magnon spectra of CrI$_3$\cite{Pengcheng18}. Even though the 
gap opening at the Dirac point is due to the presence of Dzyaloshinskii-Moriya interactions (DMI), 
the interplay between BQ exchange and DMI plays a substantial role in several features observed 
in the spin waves at different {\bf k}-points\cite{Kim11131,PhysRevX.10.011075}. In particular at the 
magnon dispersion at the K$-$M$-$K path at the Brillouin zone\cite{Kartsev:2020aa}. 
These findings provide further background on the effect of BQ 
exchange on the magnetic properties of CrI$_3$.

The shape of $M(T)$ obtained from 
the classical Monte Carlo simulations instead of 
Eq.\ref{eq:hamiltonian} shows a 
much stronger curvature than displayed 
by the measured data at low temperatures. 
To better reflect the quantum nature of the heat bath of the CrI$_3$ systems, 
we apply quantum rescaling methods\cite{Evans2015} to adjust the 
average strength of the thermal spin fluctuations within the non-Heisenberg model.
The method has previously been applied to quantitatively describe the 
temperature dependent magnetization of Fe, Co, Ni and Gd magnets. 
We extend the approach to monolayer and bulk CrI$_3$ (Fig. \ref{fig1}{\bf c}). 
Physically the temperature rescaling represents the quantum nature of the heat bath, consisting of 
discrete electron-spin and phonon-spin scattering processes. At low temperatures the spin directions 
are dominated by exchange interactions preferring ferromagnetic alignment of localized Cr spins. 
In the case of electron-spin scattering, only energetic electrons are inelastically scattered 
causing a local spin flip, while low energy electrons are elastically scattered and no spin flip occurs. 
Macroscopically this significantly reduces the average strength of the thermal spin fluctuations 
within the simulation which we approximate by applying a simple temperature rescaling of the form:
\begin{linenomath}
\begin{equation}
T_{sim} =  ( T_{exp} / T_{C} )^{1 / \alpha}
\end{equation}
\end{linenomath}
where $\alpha$ is a phenomenological rescaling exponent 
extracted from the experimental data (Fig. \ref{fig1}{\bf c}). 
The fitting assumes a simple Curie-Bloch interpolation of the form:
\begin{linenomath}
\begin{equation}
m(T) = [1-(T/T_C)^\alpha]^\beta
\label{Curie-bloch}
\end{equation}
\end{linenomath}
and is seen to fit a wide range of ferromagnetic and antiferromagnetic 
materials including the current material of interest, CrI$_3$. 
Practically our rescaling approach is 
only applicable over an ensemble average of hundreds of spins as individual 
scattering events are not directly simulated within our semi-classical method, 
but effectively introduce the quantum nature of the heat bath within a classical model. 
Nevertheless, the ability of the classical non-Heisenberg model to quantitatively 
reproduce the temperature dependent properties of bulk CrI$_3$ is remarkable. 
Figure \ref{fig1}{\bf d} highlights the difference between 
simulations with and without quantum rescaling corrections for bulk CrI$_3$. 
It is clear that the classical nature of the atomistic spin model\cite{Kuzmin05}
induced discrepancies with the measured data below \TC. 
As the spins are treated classically 
using a non-Heisenberg Hamiltonian, which follows the Boltzmann distribution, 
the curvature of the measured $M(T)$ 
deviates from the classical behaviour due to  
infinitesimal thermal fluctuations of the spins at the low temperature regime. 
These fluctuations of the magnetization have a quantum origin that are better 
represented using quantum statistics within the Bose-Einstein distribution. 

By fitting Eq. \ref{Curie-bloch} to the bulk 
experimental data with $\beta_{\rm bulk}\sim$0.25 initially 
extracted from the classical simulation 
we obtain excellent agreement between the scaled and 
measured $M(T)$ for $T<T_{C}$ at $\alpha=1.70$. 
For the monolayer data (Fig.~\ref{fig1}{\bf e}) 
we follow the same process as for the bulk, 
computing $\beta_{\mathrm{1L}} = 0.22$ from first principles and 
Monte Carlo calculations. 
Assuming that the Bloch exponent $\alpha$ is
independent of dimensionality, we find a sound agreement with the
measured data for monolayer CrI$_{\rm 3}$. 
The different $\beta$ values compared to bulk ferromagnets indicate the criticality of the
magnetization near the Curie temperature, and are not universal properties of Heisenberg 
and Ising systems in contrast with previous studies\cite{Petrovic18}. 
The Bloch exponent and therefore quantum corrections are 
sufficient to explain the different shapes of $M(T)$ curves without
the need to resort to fundamentally different models, 
for example Ising, Kitaev, or XY models (Supplementary Section 4). 
We also performed quantitative comparison between simulated and measured data for $M(T)$ 
at different magnitudes of the magnetic field B$_{\rm z}$ (Fig. \ref{fig1}{\bf f}).  
The applied B$_{\rm z}$ reduces the criticality of the magnetization close to the 
Curie temperature, and the simulations converge towards the 
experimental data mainly for temperatures below $\TC$ 
with negligible differences (less than ~1$\%$.) 
Moreover, the field dependence of the magnetization above the Curie point 
is stronger in the experiments compared to the simulations 
since we do not take into account quantum rescaling effects above $\TC$. 
The roughly double amount of B$_{\rm z}$ in the simulations to reproduce the experimental 
dependence beyond the Curie point suggests that quantum effects are still 
important as spin wave excitations or magnons may be 
present as previously observed in other magnetic materials\cite{Ernst17,Stocks05}. 
In reality an externally applied field affects the microscopic spin fluctuations 
and therefore alters the thermodynamic distribution of spins, for thin magnets it leads 
to a larger equilibrium magnetization than for the purely 
classical approach even far above $\TC$\cite{Ernst17}.

An outstanding question raised by the experiments is why 
a macroscopic magnetization or a monodomain exists 
in a 2D system after zero-field cooling. It is known that magnetic anisotropy 
overcomes the limit of the Mermin-Wagner theorem by symmetry 
breaking, but one would ordinarily expect that magnetic domains 
are stable in the system, particularly in high-anisotropy materials 
such as CrI$_3$. To investigate this we simulated the zero-field 
and field cooling processes for a large square nano-flake of 
monolayer CrI$_3$~of dimensions 0.4 $\mu$m $\times$ 0.4 $\mu$m 
using atomistic spin dynamics (see Supplementary Section 5). 
The system is thermally equilibrated above the Curie temperature and then 
linearly cooled to 0 K in a simulated time of 2 ns for different 
values of applied external magnetic field, as shown in Figure \ref{fig:2} 
and Supplementary Movies S1-S3. From the simulations 
we extract the time evolution of the spins and the formation of 
magnetic domains extracting snapshots of the spin 
configurations during the zero-field cooling process. 
For zero magnetic field shown in Fig.~\ref{fig:2}{\bf a-c} we find that the 
magnetic domains are metastable (Supplementary Movie S1) while for a 
small field of B$_{\rm z}=$10~mT 
the domains are mostly removed during the 2 ns 
cooling process (Fig.~\ref{fig:2}{\bf d-f}, Supplementary Movie S2). 
For the zero field cooling the domains persist until the end of the simulation, 
but show a continuous evolution in time at 0 K showing their metastable nature. 
As the field increases to 50 mT (Supplementary Movie S3) the domains are flushed 
out with a homogeneous magnetization being observed over the entire simulation 
area (0.4 $\mu$m $\times$ 0.4 $\mu$m) after 2 ns.  
Our observations suggest that magnetic domains are not intrinsically stable in CrI$_3$, which 
indicates a macroscopic magnetization throughout the surface even in zero-magnetic field.  
Domains as large as 0.57 $\mu$m have been observed (Supplementary Movie S1). 
Moreover, the interplay between metastability and large magnetic anisotropy could give the 
physical ingredients for the coexistence of different domain wall types in CrI$_3$. 
This effect could be intrinsic to 2D vdW magnets with wide implications for device developments and 
real applications.

Interestingly, the metastability of the domains 
prevents the wall profiles from reaching a truly ground-state configuration. 
A projection of the magnetization {\bf M} over the domain walls at 0 K and zero field (Fig.~\ref{fig:2}{\bf g}) 
shows that such unstable magnetic domains can be of several types (Fig.  \ref{fig:2}{\bf h-k}). 
This is particularly acute near the middle of the sample where quenching 
leads to a frustrated set of domains and the in-plane direction of the 
magnetizations rotates repeatedly. This effect is also observed closer to the edge 
where it is possible to observe a persistent rotation of the in-plane magnetization (Fig.\ref{fig:2}{\bf k}) 
over short lengths of the wall but extending over the entire boundary of the magnetic domains. 
For the few domain walls that can be stabilized at a specific magnetization direction 
we find that the majority of the magnetic domain walls in CrI$_3$ 
(around 97\%) are N\'eel-type (Fig.~\ref{fig:2}{\bf i}) but with some large proportion 
of a new hybrid type (Fig.~\ref{fig:2}{\bf h}) with characteristics between 
Bloch (Fig.  \ref{fig:2}{\bf j}) and N\'eel walls. 
A minor amount of domains, 
less than 3\%, stabilized at Bloch type 
over the entire system. These domains were obtained from different stochastic 
realizations of the zero field cooling simulations.

To determine whether such diverse domain walls have additional 
characteristics in monolayer CrI$_{\rm 3}$, we project the total magnetization at the wall over 
in-plane (M$_{\rm x}$, M$_{\rm y}$) and out-of-plane (M$_{\rm z}$) components (Fig.~\ref{fig:3}{\bf a}, {\bf c} {\bf e}). While M$_{\rm z}$ through Bloch, hybrid and N\'eel domains  
does not change appreciably, both M$_{\rm x}$ and M$_{\rm y}$ show different 
behaviour characterizing a specific kind of domain wall with its specific 
spin orientations (Fig.~\ref{fig:3}{\bf b}, {\bf d} {\bf f}). 
It is noteworthy that the hybrid domains have a different chirality for the in-plane 
moments relative to Bloch and N\'eel with a sizable component along of the $y$ axis as the spins transition 
from one domain to another (Fig.~\ref{fig:3}{\bf c},{\bf d} and Fig.~\ref{fig:2}{\bf h}). 
We can extract the domain wall width $\delta$ by fitting the different components of the 
magnetization (M$_x$, M$_y$, M$_z$) to a standard equation profile of the form: 
%
\begin{linenomath}
\begin{equation}
M_{}(r) = {\rm tanh} (\pi (r-r_0)/\delta) 
\end{equation}
\end{linenomath}
where $r_0$ is the domain wall position at a specific orientation ($x$, $y$, $z$).  
All types of wall have a very narrow domain wall width of 
around $\delta \sim 3.8-4.8$ nm (Fig.~\ref{fig:3}{\bf a},{\bf c},{\bf e}). 
Such small domain walls are typically only seen in permanent magnetic 
materials due to the exceptionally high magnetic anisotropy\cite{Hubert-book}. 
For such materials the magnetic domains are stabilized in a zero-remanence state after zero-field 
cooling due to the long-ranged dipole-dipole interactions which 
are also taken into account in our calculations (Supplementary Section S9). 
Nevertheless we find that this is not the case for monolayer CrI$_3$ which suggests 
that this material reunites features from a soft-magnet (e.g. easy movement of domain walls, 
small area hysteresis loop) and a hard-magnet (e.g. relative high 
magnetocrystalline anisotropy, narrow domain walls).  

The variety of domain-walls observed in \criii can be directly related to the 
magnetic stability of the layer\cite{Hubert-book}. For magnetic materials 
with strong uniaxial anisotropy, the equilibrium state is normally reached 
beyond the field cooling process\cite{Wulferding:2017aa}. 
Even though no thermal energy would be available at such limit, 
the spins would still evolve to stabilize the ground-state via the 
minimization of other contributions of the total 
energy, e.g. exchange, anisotropy. This process can be 
observed in Figure \ref{fig4} for the time-evolution 
of one of the spin dynamics of monolayer CrI$_3$ once the system 
had achieved 0 K within 2.0 ns at zero field. There is a continuous 
modification of the domain-wall profiles through all components of the 
magnetization (M$_{x,y,z}$) over time. The variations on M$_{z}$ 
across the magnetic domains (Fig. \ref{fig4}{\bf a-b}) tend to 
be smooth without sudden changes differently to those observed 
along the in-plane components (Fig. \ref{fig4}{\bf c-d}). For them, several 
peaks appeared and vanished on a time scale of few tenths of nanoseconds indicating 
the stochastic nature of the spin-fluctuations in the system. Indeed, we observed 
such random fluctuations of M$_{x,y}$ even beyond 20 ns which suggest 
that the system may be intrinsically not a local minimum but rather at a
flat energy landscape.    
We can extract some qualitative information about the magnetic behaviour of the domains in \criii 
regarding stability and domain size using magnetic force 
microscopy (MFM) experiments (see Supplementary Section S10 for details). 
Supplementary Figures S18-S19 shows a zero-field cooled CrI$_3$ thick flake (0.04 $\mu$m) 
with lateral dimensions of approximately 4 $\mu$m $\times$ 2 $\mu$m at 4.2 K where 
magnetic domains of about 2 $\mu$m persist to the base temperature (Supplementary Fig. S19{\bf a-b}).  
Even though the measurements were undertaken at a sample area one order of magnitude 
larger than that utilized in the simulations (e.g. 0.16 $\mu$m$^{\rm 2}$), 
the magnetic domains formed in both theory and experiments still 
keep the same scale relative to the domain size created. 
This indicates that mostly a monodomain is created over 
the entire surface as suggested by the theoretical results. 
Moreover, the topography 
of the domains extracted from frequency shift profiles (Suppl. Fig. S19{\bf f-i}) 
clearly shows sharp domain walls (e.g. smaller than 20 nm) 
but the resolution limitation of the MFM technique 
(50 nm diameter of an average Cr coated MFM tip) prevents 
the direct comparison with the sub-10 nm prediction extracted from theory.
Similar limitation ($\sim$40-60 nm) was also observed in recent 
measurements using a scanning single-spin magnetometry with a 
Nitrogen-Vacancy (NV) centre spin in the tip of an atomic force 
microscope\cite{Maletinsky19} and a magnetic-circular dichroism 
technique\cite{Zhong:2020aa}. 
This indicates that further development 
on the experimental side is needed to further validate the 
simulation results. It is worth mentioning that even though 
the time-scale used in the spin dynamics 
spans around 20 ns, it reproduces accurately the domain structure 
obtained via MFM over a few hours scan process. The images recorded 
in Supplementary Figure S19 may be considered as the final magnetic state 
as several electronic and spin interactions take place. The early stages 
which determined the magnetic ordering can be extracted 
from the micro-magnetic simulations as a sound agreement is obtained with the recorded 
MFM images. 
In addition, atomistic simulations undertake in 
bulk CrI$_3$ (Supplementary Figure S20) 
support the picture that magnetic 
domains are more stable in bulk (Supplementary Figures S18-S19) 
than in monolayer due to the additional interlayer interactions 
driven by vdW forces and spin exchange. 
Thus, the meta-stability 
of the magnetic domains seems to be more present in the 
lower dimensionality of single sheets.

%
The magnetism of 2D materials at the limit of one or few layers is still at its early 
stages with rich phenomena yet to be explored. The demonstration here of quantum 
effects in CrI$_3$ together with its non-Heisenberg character 
due to higher-order exchange interactions should motivate 
significant future studies to understand both the mechanism of the 
computed enhancement of biquadratic interactions and to confirm 
that such effect may be general to several families of 2D vdW magnets. 
In addition, the metastability of the magnetic domains in CrI$_3$ induces a 
homogeneous magnetization or even a single domain 
over the entire surface.  This behavior associated with the out-of-plane anisotropy 
and the higher coercivity of CrI$_3$ indicates a potential 
magnetic media for perpendicular recording. 
It is still unclear however which kind of domain motion can be foreseen 
in such thin layered compound, and how the coexistence of different domain types
can affect device architectures. 
This suggests new routes for magnetic-domain engineering 
at the atomic limit.

\section*{Supplementary Materials}
\label{sec:org3881bef}

Materials and Methods. 
\\
Supplementary sections S1 to S11, movies S1, S2, S3 and Figs. S1 to S20.


\subsubsection{Data Availability}

The data that support the findings of this study 
are available within the paper and its Supplementary Information.  

\subsubsection{Competing interests}
The Authors declare no conflict of interests.

\subsubsection{Acknowledgments}
RFLE gratefully acknowledges the financial support of the Engineering and Physical Sciences 
Research Council (Grant No. EPSRC EP/P022006/1) and the use of the VIKING Cluster, 
which is a high performance compute facility provided by the University of York. 
This work was enabled by code enhancements to the VAMPIRE software 
implemented under the embedded CSE programme (ecse0709) and (ecse1307) 
of the ARCHER UK National Supercomputing Service. 
EJGS acknowledges computational resources through the 
UK Materials and Molecular Modelling Hub for access to THOMAS supercluster, 
which is partially funded by EPSRC (EP/P020194/1); CIRRUS Tier-2 HPC 
Service (ec131 Cirrus Project) at EPCC (http://www.cirrus.ac.uk) funded 
by the University of Edinburgh and EPSRC (EP/P020267/1); 
ARCHER UK National Supercomputing Service (http://www.archer.ac.uk) via 
Project d429. EJGS acknowledges the 
EPSRC Early Career Fellowship (EP/T021578/1) and 
the University of Edinburgh for funding support. 
ENM acknowledges the European Research Council 
(ERC) under the Horizon 2020 research and innovation programme 
(ERC StG, grant agreement No. 803092).

\subsubsection{Author Contributions}
EJGS conceived the idea and supervised the project. 
MA and DAW performed ab initio 
and Monte Carlo (MC) simulations under the supervision of EJGS. 
RFLE implemented the biquadratic exchange interactions in 
VAMPIRE, and also undertook MC simulations. SJ implemented 
the atomistic dipole-dipole solver to verify the computed domain wall profiles.
EVM and SMV fabricated and characterized the samples. 
IJVM, WK, KSN measured the samples using SQUID.  
EVM, SMV, EC performed the MFM measurements. 
EVM, IJVM, IVG, KSN analyzed the data and contributed 
to the discussions. EJGS wrote the paper with inputs from all authors. 
All authors contributed to this work, 
read the manuscript, discussed the results, and agreed 
to the contents of the manuscript.

\section*{References and Notes}
\bibliography{ref_bq_v020119}


\subsubsection*{Figure captions}

\begin{figure}[htbp]
\centering
\caption{\label{fig1}\textbf{Quantum rescaling corrections.} 
{\bf a-b,} Comparison of measured and calculated temperature dependent 
magnetization ($M/M_{s}$) of bulk and monolayer (1L) CrI$_3$, 
respectively, using Ising and non-Heisenberg models (Eq. \ref{eq:hamiltonian}). 
In both cases, the Ising model leads to large over-estimation of the 
Curie temperature (\TC) relative to the experiments. Even 
with the inclusion of biquadratic exchange into the description of the 
spin interactions, some deviations relative to the experiments are observed at low-temperatures. 
{\bf c,} Plot of the effective simulation (spin) temperature 
against the comparable experimental temperature of the 
environment for different values of the phenomenological 
rescaling exponent $\alpha$. For $\alpha= 1$ the two 
temperatures are equal and represents the usual situation 
for a classical Heisenberg magnet. For increasing values of $\alpha$ 
the effective spin temperature is reduced due to the quantum 
nature of the heat bath reducing the spin fluctuations. 
The value of $\alpha = 1.70179$ extracted from the measured data 
for bulk CrI$_3$ is shown for comparative purposes.
{\bf d,} Detailed comparison of the classical non-Heisenberg simulation 
and the experimental data for bulk CrI$_3$ at a magnetic field of B$_{\rm z}=1.0$ T. 
The linear behaviour of the magnetization at low temperatures is a 
well-known deficiency of a classical model. 
Applying quantum rescaling to include the quantum nature of 
the heat bath gives a quantitative agreement with the experimental data 
at the low temperature regime. At elevated temperatures, the differences arise due 
to the presence of an external magnetic field which resulted in values of \TC 
of 69 K from simulations and 63 K from experiments. 
%
{\bf e,} Magnetization as a function of the reduced temperature ($T/\TC$) 
for monolayer CrI$_3$ comparing classical and quantum rescaling-corrected 
simulations with the experimental data. 
The data is plotted normalized to T$_{\rm c}$ due to the 
small difference between measured and calculated Curie 
temperatures to enable a direct comparison of the top 
of the magnetization curve. The fitted line to the experimental 
data uses the computed value of $\beta= 0.22 \pm 0.004$ from 
the classical simulation and assumed temperature 
rescaling exponent $\alpha =$ 1.70 fitted from the 
bulk experimental data. 
{\bf f,} Comparative simulations of the temperature dependent magnetization 
for bulk \criii in different applied magnetic fields including temperature 
rescaling and normalized to \TC. The simulations 
show a sound agreement with the experimental data at 
temperatures less than \TC, while above \TC the 
apparent paramagnetic susceptibility is lower in the simulations 
due to an absence of quantum effects above 
the Curie temperature.}
\end{figure}

\begin{figure}[htbp]
\centering
\caption{\label{fig:2} {\bf Theory-predicted magnetic domains in monolayer CrI$_3$.} 
{\bf a-c,} and {\bf d-f,} Magnetic domain configurations obtained during field cooling 
at 0 mT and 10 mT, respectively. 
Bright and dark areas at T$\leq$ 16 K correspond to spins 
pointed along the easy-axis  
in different spin polarizations (e.g. up or down). 
Purple and mixed colours correspond to different spin 
orientations either before stabilization of the domains 
at T$\geq$ 16 K, or at the domain walls at T$\leq$ 16 K. 
As the system cools down the magnetic domains coalesce 
to form a circular shape to minimize the domain wall energy. 
Domains anti-parallel to the field direction are unstable, 
and eventually reverse leaving a saturated domain state at low temperatures.
{\bf g,} Analysis of the domain walls from {\bf c,} at different parts of the crystals 
undertaking an in-plane projection of the 
magnetization ${\vec{\rm M}}$ according to its colour 
orientation at the domain walls. A coexistence of several domain wall types is 
observed through {\bf h,} N\'eel, {\bf i,} Bloch, {\bf j,} hybrid, and {\bf k,} 
mixed domain walls. A continuous rotation of the spins is observed 
in the hybrid domains which extends from few tens of \AA~up to few nm's. 
}
\end{figure}

\begin{figure}[htbp]
\centering
\caption{\label{fig:3}\textbf{Simulated hybrid domain-walls.}  Plot of the 
domain wall profiles for metastable domain walls at T = 0 K for two different 
stochastic realizations in monolayer CrI$_3$. Three characteristic shapes are seen: 
{\bf a-b,} Bloch type, where the in-plane magnetization (M$_{\rm x}$, M$_{\rm y}$) 
is parallel to the domain wall. 
{\bf c-d,} Hybrid type, where the in-plane magnetization is between N\'eel and Bloch 
type and lies at some angle to the wall direction. 
{\bf e-f,} N\'eel-N\'eel-type where the in-plane magnetization is perpendicular to the domain wall. 
The schematics in the right show a visualization of the individual 
spin directions in the domain wall. Note that the different sign 
of the $x$ and $y$ components indicates a different domain wall 
chirality. The out of plane magnetization (M$_{\rm z}$) does not show appreciable
modifications over the three domains observed. 
The calculated domain wall width is in the range of 3.8-4.8 nm. 
Such narrow widths are typically only found in 
permanent magnets such as L10-FePt nanoparticles 
or Nd$_2$Fe$_{14}$B crystals\cite{Hubert-book}.  
} 
\end{figure}

\begin{figure}[htbp]
\centering
\caption{\label{fig4}\textbf{Modelling of spin fluctuations across domains.} {\bf a,} Snapshot of a spin dynamics of monolayer CrI$_3$ obtained through zero-field cooling down process after 2.00 ns and reaching 0 K . 
The magnetization along of the easy-axis (M$_z$) is displayed showing 
the domain formation. Bright (dark) areas correspond to  M$_z=\pm 1$, respectively. 
A path (dashed line) connecting three points A, B and C at the boundary 
between different magnetic domains is showed. 
{\bf b-d,} Variations of M$_z$ and the out-of-plane 
components of the magnetization (M$_x$, M$_y$), respectively,  
along of A$-$B$-$C at different times (2.40-3.80 ns) after 0 K is obtained. 
The shaded areas determine the regions considered along the path. 
}
\end{figure}




\setcounter{figure}{0}

\begin{figure}[htbp]
\centering
\includegraphics[width=1.01\linewidth]{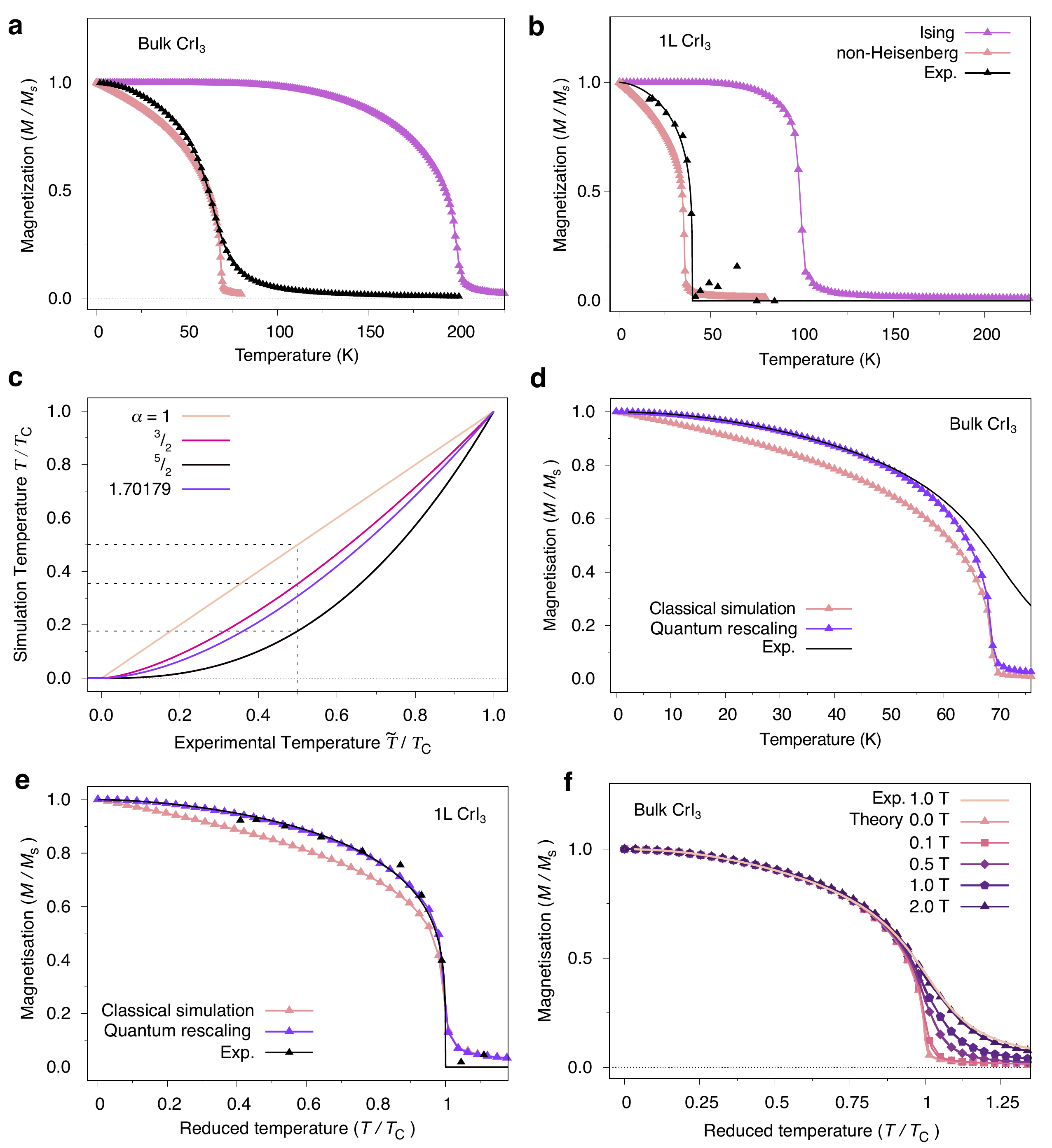}
\caption{}
\end{figure}

\begin{figure}[htbp]
\centering
\includegraphics[width=0.93\linewidth]{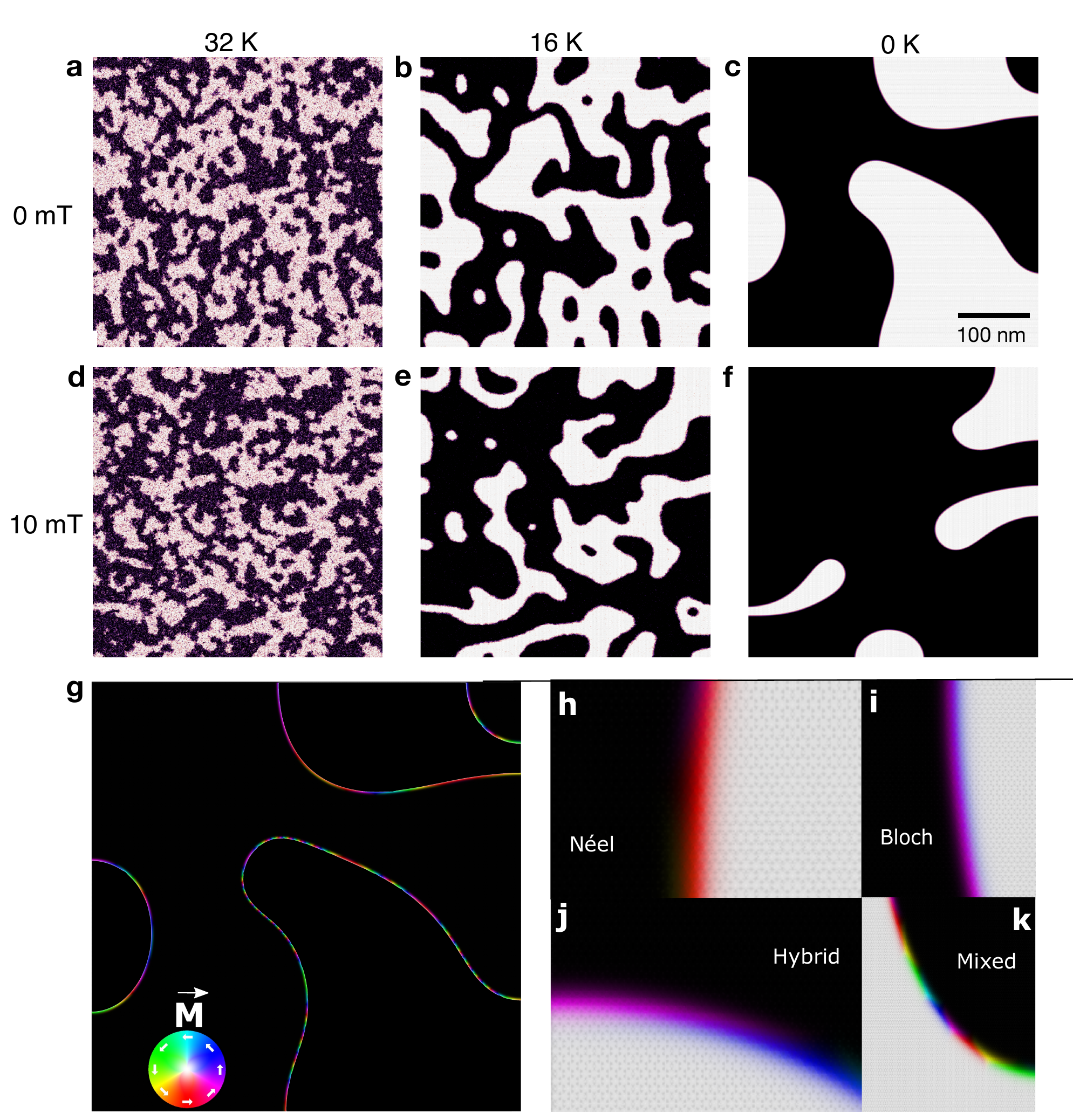}
\caption{}
\end{figure}

\begin{figure}[htbp]
\centering
\includegraphics[width=1.10\linewidth]{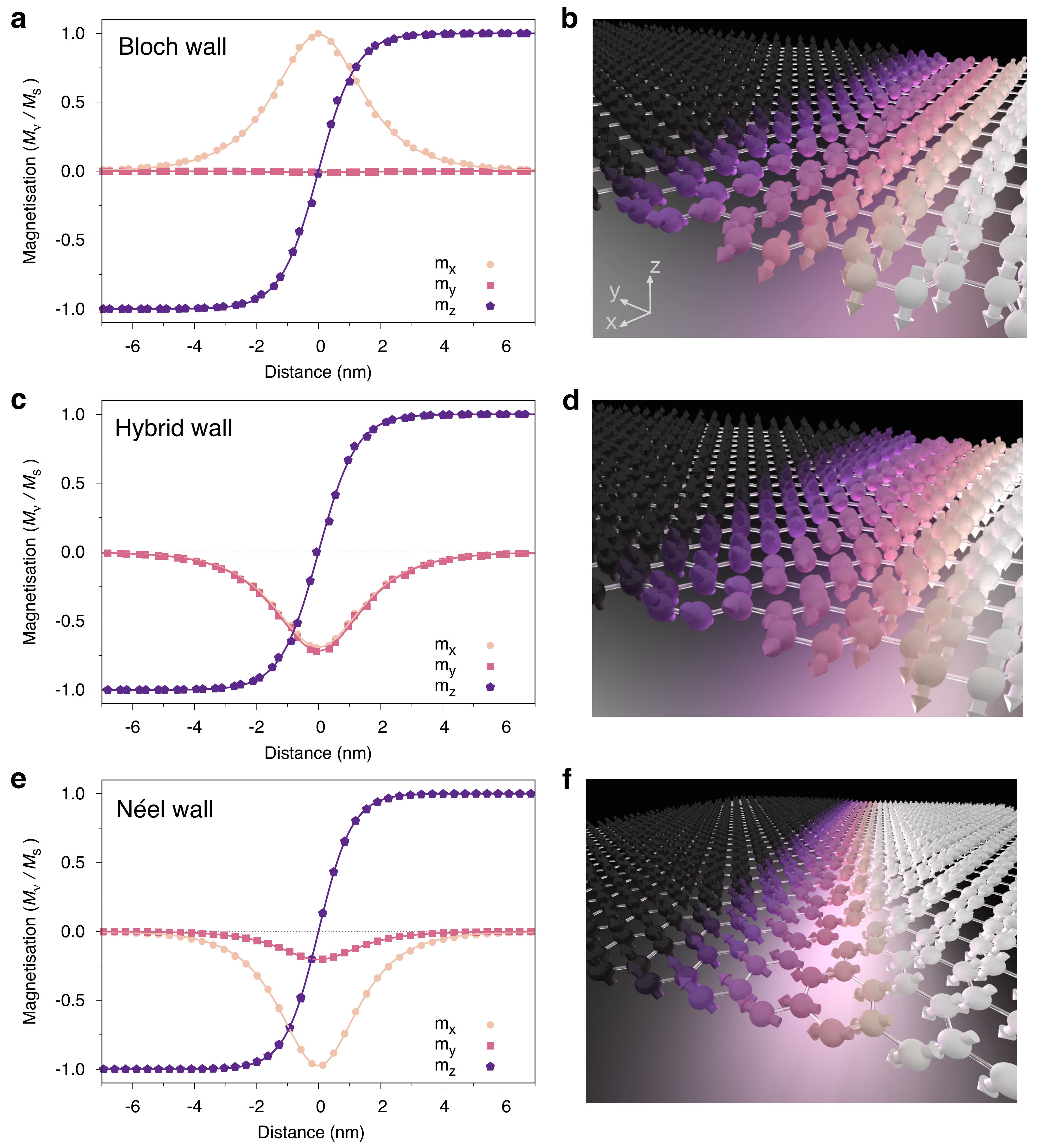}
\caption{}
\end{figure}

\begin{figure}[htbp]
\centering
\includegraphics[width=1.10\linewidth]{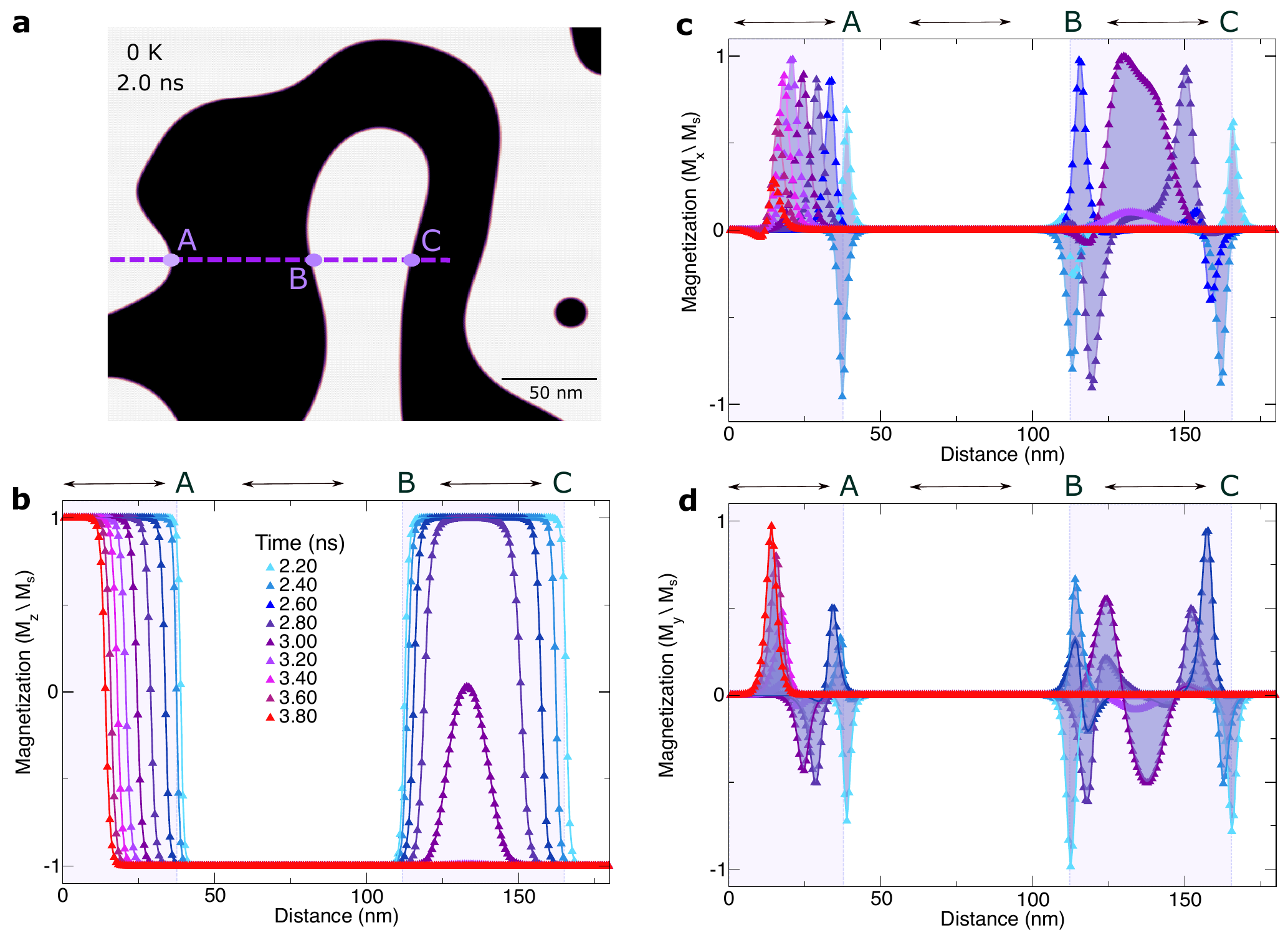}
\caption{}
\end{figure}


\clearpage


\end{document}